\documentclass[runningheads,fleqn]{cl2emult}

\usepackage{makeidx}  
\usepackage{graphicx} 
\usepackage{subeqnar} 
\usepackage{multicol} 
\usepackage{cropmark} 
\usepackage{comp}     

\usepackage{amsmath}
\usepackage{amssymb}
\usepackage{amscd}
\usepackage{latexsym}
\usepackage{epsfig}

\newtheorem{defi}{Definition}
\newtheorem{thm}[defi]{Theorem}

\newtheorem{rem}[defi]{Remark}

\newtheorem{conj}[defi]{Conjecture}

\newtheorem{exempel}[defi]{Example}

\newlength{\blank}
\begin{document}

\newcommand{\1}{{\mathbb{1}}}
\newcommand{\bra}[1]{{\langle{#1}|}}
\newcommand{\ket}[1]{{|{#1}\rangle}}
\newcommand{\Z}{{\mathbb{Z}}}

\title*{Quantum Finite State Transducers}

\toctitle{Quantum Finite State Transducers}

\titlerunning{Quantum Finite State Transducers}

\author{R\= usi\c n\v s Freivalds\inst{1}
\and Andreas Winter\inst{2}}

\authorrunning{R\= usi\c n\v s Freivalds and Andreas Winter}

\institute{Institute of Mathematics and Computer Science, University of Latvia, Rai\c na bulv\= aris 29, LV--1459, Riga, Latvia. Email: \texttt{rusins@paul.cclu.lv}.
\and 
Department of Computer Science, University of Bristol, Merchant Venturers Building, Woodland Road, Bristol BS8 1UB, United Kingdom. Email: \texttt{winter@cs.bris.ac.uk}.}

\maketitle              


\begin{abstract}
  We introduce \emph{quantum finite state transducers (qfst)}, and
  study the class of relations which they compute. It turns
  out that they share many features with probabilistic finite
  state transducers, especially regarding undecidability of
  emptiness (at least for low probability of success). However,
  like their `little brothers', the quantum finite automata,
  the power of qfst is incomparable to that of their probabilistic
  counterpart. This we show by discussing a number of characteristic
  examples.
\end{abstract}


\section{Introduction and definitions}
\label{sec:intro}
\label{sec:defi}
The issue of this work is to introduce and to study the computational model
of quantum finite state transducers. These can be understood as finite automata
with the addition of an output tape
which compute a relation between strings, instead of a decision (which we read
as a binary valued function). After the necessary definitions, the relation to
quantum finite automata is clarified (section~\ref{sec:qfa:qfst}), then
decidability questions are addressed (section~\ref{sec:empty}):
it is shown that emptiness of the
computed relation is undecidable both for quantum and probabilistic transducers.
However, the membership problem for a specific output is decidable.
Next, the relation between deterministic and probabilistic transducers is explored
(section~\ref{sec:det:prob}), and in section~\ref{sec:qfst}
quantum and probabilistic transducers are compared.
\par
We feel our extension of quantum automata studies to this new model
justified by the following quote from D. Scott~\cite{scott}:
\begin{quote}
  {\it
    `The author (along with many other people) has come recently
  to the conclusion that the functions computed by the various
  machines are more important -- or at least more basic -- than
  the sets accepted by these devices. (...) In fact by putting the
  functions first, the relationship between various classes of sets
  becomes much clearer'.}
\end{quote}
\par
We start be reviewing the concept of probabilistic finite state transducer.
For a finite set $X$ we denote by $X^*$ the set of all finite strings formed
from $X$, the empty string is denoted $\epsilon$.

\begin{defi}
  \label{defi:pfst}
  A \emph{probabilistic finite state transducer (pfst)} is a tuple
  $T=(Q,\Sigma_1,\Sigma_2,V,f,q_0,Q_{\rm acc},Q_{\rm rej})$, where $Q$ is a finite
  set of states, $\Sigma_1,\Sigma_2$ is the input/output
  alphabet, $q_0\in Q$ is the initial state, and
  $Q_{\rm acc},Q_{\rm rej}\subset Q$ are (disjoint) sets of
  accepting and rejecting states, respectively. (The
  other states, forming set $Q_{\rm non}$, are called non--halting).
  The transition function $V:\Sigma_1\times Q\rightarrow Q$ is
  such that for all $a\in\Sigma_1$ the matrix $(V_a)_{qp}$
  is stochastic, and $f_a:Q\rightarrow\Sigma_2^*$ is the
  output function. If all matrix entries are either $0$ or $1$
  the machine is called a \emph{deterministic finite state
  transducer (dfst)}.
\end{defi}
The meaning of this definition is that, being in state $q$, and reading
input symbol $a$, the transducer prints $f_a(q)$ on the output
tape, and changes to state $p$ with probability $(V_a)_{qp}$, moving
input and output head to the right. After each such step,
if the machine is found in a halting state, the computation
stops, accepting or rejecting the input, respectively.
\par
To capture this formally, we introduce the \emph{total state}
of the machine, which is an element
$$(P_{\rm NON},P_{\rm ACC}, p_{\rm rej})\in
\ell^1(Q\times\Sigma_2^*)\oplus \ell^1(\Sigma_2^*)\oplus \ell^1(\{{\rm REJ}\}),$$
with the natural norm
$$\|(P_{\rm NON},P_{\rm ACC},p_{\rm rej})\|=
                             \|P_{\rm NON}\|_1+\|P_{\rm ACC}\|_1+|p_{\rm rej}|.$$
At the beginning, the total state is $((q_0,\epsilon),{\bf 0},0)$
(where we identify an element of $Q\times\Sigma_2^*$ with its characteristic
function). The computation is represented by the (linear extensions
of the) transformations
$$T_a:((q,w),P_{\rm ACC},p_{\rm rej})\mapsto
     \left(\left(\sum_{p\in Q_{\rm non}} (V_a)_{qp}p,w f_a(q)\right),
                                          P_{\rm ACC}',p_{\rm rej}'\right),$$
of the total state, for $a\in\Sigma_1$, with
\begin{equation*}
  P_{\rm ACC}'(x)=\begin{cases}
                    P_{\rm ACC}(x)+\sum_{p\in Q_{\rm acc}} (V_a)_{qp}
                                                        & \text{ if } x=w f_a(q),\\
                    P_{\rm ACC}(x)                      & \text{ else},
                  \end{cases}
\end{equation*}
and $p_{\rm rej}'=p_{\rm rej}+\sum_{p\in Q_{\rm rej}} (V_a)_{qp}$.
\par
For a string $x_1\ldots x_n$ the map $T_x$ is just the concatenation of the
$T_{x_i}$. Observe that all the $T_a$ conserve the probability.
\par
Implicitely, we add initial and end marker symbols ($\ddag,\$ $)
at the input, with additional stochastic matrices $V_\ddag$ and
$V_\$ $, executed only at the very beginning, and at the very
end. We assume that $V_\$ $ puts no probability outside
$Q_{\rm acc}\cup Q_{\rm rej}$.
\par
By virtue of the computation, to each input string $v\in\Sigma_1^*$
there corresponds a probability distribution $T(\cdot|v)$
on the set $\Sigma_2^*\cup\{{\rm REJ}\}$:
$$T({\rm REJ}|v):=T_{\ddag v \$}((q_0,\epsilon),{\bf 0},0)[{\rm REJ}]$$
is the probability to reject the input $v$, whereas
$$T(w|v):=T_{\ddag v \$}((q_0,\epsilon),{\bf 0},0)[w]$$
is the probability to accept, after having produced the
output $w$.
\begin{defi}
  \label{defi:compute}
  Let ${\cal R}\subset\Sigma_1^*\times\Sigma_2^*$.
  \par
  For $\alpha>1/2$ we say that $T$ \emph{computes the relation ${\cal R}$
    with probability $\alpha$} if for all $v$, whenever $(v,w)\in{\cal R}$,
  then $T(w|v)\geq\alpha$, and whenever $(v,w)\not\in{\cal R}$,
  then $T(w|v)\leq 1-\alpha$
  \par
  For $0<\alpha<1$ we say that $T$ \emph{computes the relation ${\cal R}$
    with isolated cutpoint $\alpha$} if there exists $\varepsilon>0$ such
  that for all $v$, whenever $(v,w)\in{\cal R}$, then
  $T(w|v)\geq\alpha+\varepsilon$, but whenever
  $(v,w)\not\in{\cal R}$, then $T(w|v)\leq\alpha-\varepsilon$.
\end{defi}
The following definition is modelled after the ones for pfst
for quantum finite state automata~\cite{kondacs:watrous}:
\begin{defi}
  \label{defi:qfst}
  A \emph{quantum finite state transducer (qfst)} is a tuple
  $T=(Q,\Sigma_1,\\ \Sigma_2, V,f,q_0,Q_{\rm acc},Q_{\rm rej})$,
  where $Q$ is a finite
  set of states, $\Sigma_1,\Sigma_2$ is the input/output
  alphabet, $q_0\in Q$ is the initial state, and
  $Q_{\rm acc},Q_{\rm rej}\subset Q$ are (disjoint) sets of
  accepting and rejecting states, respectively.
  The transition function $V:\Sigma_1\times Q\rightarrow Q$ is
  such that for all $a\in\Sigma_1$ the matrix $(V_a)_{qp}$
  is unitary, and $f_a:Q\rightarrow\Sigma_2^*$ is the
  output function.
\end{defi}
Like before, implicitely matrices $V_\ddag$ and $V_\$ $ are assumed,
$V_\$ $ carrying no amplitude from $Q_{\rm non}$ to outside
$Q_{\rm acc}\cup Q_{\rm rej}$.
The computation proceeds as follows: being in state $q$, and reading
$a$, the machine prints $f_a(q)$ on the output tape, and moves
to the superposition $V_a\ket{q}=\sum_p (V_a)_{qp}\ket{p}$ of internal states.
Then a measurement of the orthogonal
decomposition $E_{\rm non}\oplus E_{\rm acc}\oplus E_{\rm rej}$
(with the subspaces $E_i={\rm span}\ Q_i\subset \ell^2(Q)$, which we identify
with their respective projections) is performed, stopping
the computation with accepting the input on the second outcome
(while observing the output),
with rejecting it on the third.
\par
Here, too, we define total states: these are elements
$$(\ket{\psi_{\rm NON}},P_{\rm ACC},p_{\rm rej})\in
\ell^2(Q\times\Sigma_2^*)\oplus \ell^1(\Sigma_2^*)\oplus \ell^1(\{{\rm REJ}\}),$$
with norm
$$\|(\ket{\psi_{\rm NON}},P_{\rm ACC},p_{\rm rej})\|=
                  \|\ket{\psi_{\rm NON}}\|_2+\|P_{\rm ACC}\|_1+|p_{\rm rej}|.$$
At the beginning the total state is $(\ket{q_0}\otimes\ket{\epsilon},{\bf 0},0)$,
the total state transformations, for
$$\ket{\psi}=\sum_{q\in Q} \ket{q}\otimes\ket{\omega_q},\qquad\text{with }
  \ket{\omega_q}=\sum_{w\in\Sigma_2^*} \alpha_{qw}\ket{w},$$
are (for $a\in\Sigma_1$)
$$T_a:(\ket{\psi},P_{\rm ACC},p_{\rm rej})\mapsto
          \left(E_{\rm non}\sum_q V_a\ket{q}\otimes\ket{\omega_q f_a(q)},
                P_{\rm ACC}',
                p_{\rm rej}'\right),$$
where $\ket{\omega_q f_a(q)}=\sum_w \alpha_{qw}\ket{wf_a(q)}$, and
\begin{align*}
  P_{\rm ACC}'(x) &=P_{\rm ACC}(x)
                    +\left\|E_{\rm acc}\sum_{q,w\text{ s.t. }x=wf_a(q)} 
                                          \alpha_{qw}V_a\ket{q}\right\|_2^2\ ,\\
  p_{\rm rej}'    &=p_{\rm rej}
                    +\left\|E_{\rm rej}
                      \sum_q V_a\ket{q}\otimes\ket{\omega_q f_a(q)}\right\|_2^2\ .
\end{align*}
Observe that the $T_a$ do not exactly preserve the norm, but that there
is a constant $\gamma$ such that $\|T_a(X)\|\leq\gamma\|X\|$ for any total
state $X$.
Quite straightforwardly, the distributions $T(\cdot|v)$ are defined, and so are the
concepts of computation with probability $\alpha$ or with isolated
cutpoint $\alpha$.
\par
Observe also that we defined our model in closest possible analogy to
quantum finite automata~\cite{kondacs:watrous}. This is of course to
be able to compare qfst to the latter. In principle however other
definitions are conceivable, e.g. a mixed state computation where
the $T_a$ are any completely positive, trace preserving, linear maps
(the same of course applies to quantum finite automata!).
We defer the study of such a model to another occasion.
\par
Notice the physical benefits of having the output tape: whereas for
finite automata a superposition of states means that the amplitudes
of the various transitions are to be added, this is no longer true
for transducers if we face a superposition of states \emph{with different
output tape content}. I.e. the entanglement of the internal state with
the output may prohibit certain interferences. This will be a crucial
feature in some of our later constructions.

\section{Quantum Finite Automata and Quantum Transducers}
\label{sec:qfa:qfst}
The definition of qfst is taylored in such a way that by excluding
the output tape and the output function, we get a quantum finite
automaton. One, however, with distinct acceptance and rejection
properties, as compared to the qfst.
Nevertheless, the decision capabilities
of qfst equal those of quantum finite automata:
\begin{thm}
  \label{thm:qfa-c-qfst}
  A language $L$ is accepted by a 1--way quantum finite automaton
  with probability bounded away from 1/2 if and only if the relation
  $L\times\{0\}\cup\overline{L}\times\{1\}$ is computed with isolated
  cutpoint.
\end{thm}
{\it Proof}:
  First observe that for finite automata (probabilistic and quantum), recognizability
  with an isolated cutpoint is equivalent to recognizability with probability
  bounded away from $1/2$ (by ``shifting the cutpoint'': just add
  in the $\ddag$--step possibilities to accept or reject right away with
  certain probabilities).
  We have to exhibit two constructions:
  \par
  Let there be given a quantum finite automaton. We may assume that it is such
  that $V_\$ $ is a permutation on $Q$.
  \par
  This can be forced by duplicating
  each $q\in Q_{\rm acc}\cup Q_{\rm rej}$ by a new state $q'$, and modifying the transition
  function as follows: denote by $\sigma$ the map interchanging $q$ with $q'$
  for $q\not\in Q_{\rm non}$, and being the identity on $q,q'$ for $q\in Q_{\rm non}$.
  Define a unitary $U$ such that for $q\in Q_{\rm non}$
  $$U\ket{q}=\sum_p (V_\$)_{qp}\ket{\sigma p},$$
  and $U\ket{q}=\ket{q}$ for $q\in Q_{\rm acc}\cup Q_{\rm rej}$.
  Now let
  $$V_\ddag' :=UV_\ddag,\qquad
    V_\$'    :=\sigma,\qquad
    V_a'     :=UV_a U^{-1}.$$
  It is easily checked that this automaton behaves exactly like the 
  initial one.
  \par
  Construct a qfst as follows: its states are
  $Q\cup\widehat{Q}$,
  with $\widehat{Q}=\{\hat{q}:q\in Q_{\rm acc}\cup Q_{\rm rej}\}$ being
  the accepting states, and no rejecting states.
  Let the transition function be $W$ with
  \begin{align*}
    W_a\ket{q} &=V_a\ket{q}\text{ for }q\in Q_{\rm non},\text{ but}\\
    W_a\ket{q} &=\ket{\hat{q}}\text{ for }q\in Q_{\rm acc}\cup Q_{\rm rej}.
  \end{align*}
  Since $V_\$ $ is the permutation $\sigma$ on $Q$, we may define
  \begin{equation*}
    W_\$\ket{q}=\begin{cases}
                  \ket{\widehat{\sigma q}} & \text{for }\sigma q\in Q_{\rm acc}\cup Q_{\rm rej}, \\
                  \ket{\sigma q}           & \text{for }\sigma q\in Q_{\rm non}.
                \end{cases}
  \end{equation*}
  Finally, let the output function be (for $q\in Q$)
  \begin{equation*}
    \begin{array}{lcr}
      {
        f_a(q)=\begin{cases}
                 0        & \text{for }q\in Q_{\rm acc},\\
                 1        & \text{for }q\in Q_{\rm rej},
               \end{cases}
      } & {\phantom{===}} &
      {
      f_\$(q)=\begin{cases}
                0        & \text{for }\sigma q\in Q_{\rm acc},\\
                1        & \text{for }\sigma q\in Q_{\rm rej},
              \end{cases}
      }
    \end{array}
  \end{equation*}
  and $\epsilon$ in all other cases. It can be checked that it
  behaves in the desired way.
  \par
  Given a qfst, construct a quantum finite automaton as follows:
  its states are
  $Q\times\Sigma_2^{\leq t}$, where the second
  component represents the tape content up to
  $t=1+\max_{a,q} |f_a(q)|$ many symbols. Initial state is $(q_0,\epsilon)$.
  Observe that by definition of the $T_a$ amplitude that once is shifted onto
  output tapes of length larger than $1$ is never recovered for smaller
  lengths. Hence we may as well cut such branches by immediate rejection:
  the states in $Q\times\Sigma_2^{\geq 2}$ are all rejecting, and
  so are $(Q_{\rm acc}\cup Q_{\rm rej})\times\{1\}$.
  The accepting states are $Q_{\rm acc}\times\{0\}$.
  \par
  The transition function is partially defined by
  $$W_a\ket{q,x}:=\sum_{p\in Q} (V_a)_{qp}\ket{p,xf_a(q)},\quad x\in\Sigma\cup\{\epsilon\},$$
  (for $a=\$ $ this is followed by mapping $\ket{p,\epsilon}$
  to a rejecting state, while leaving the other halting states
  alone), i.e. the automaton performs like the qfst on the
  elements of $Q$, and uses
  the second component to simulate the output tape. We think of $W_a$
  being extended in an arbitary way to a unitary map.
  One can check that this construction behaves in the desired way.
\qed\par

\section{Decidability questions}
\label{sec:empty}
As is well known, the emptiness problem for the language accepted
by a deterministic (or nondeterministic) finite automaton is decidable.
Since the languages accepted by probabilistic
and quantum finite automata with bounded error
are regular~\cite{rabin,kondacs:watrous},
these problems are decidable, too.
\par
For finite state transducers the situation is more complicated:
In~\cite{whoever} it is shown that the emptiness problem
for deterministic and nondeterministic fst is decidable.
In contrast we have
\begin{thm}
  \label{thm:empty1}
  The emptiness problem for pfst computing a relation with
  probability $2/3$ is undecidable.
  \par
  Likewise, the emptiness problem for qfst computing a relation with
  probability $2/3$ is undecidable.
\end{thm}
{\it Proof}:
By reduction to the Post Correspondence Problem: let an instance
$(v_1,\ldots,v_k)$, $(w_1,\ldots,w_k)$ of PCP be given (i.e.
$v_i,w_i\in\Sigma^+$). It is to be decided whether there exists
a sequence $i_1,\ldots,i_n$ ($n>0$) such that
$$v_{i_1}\cdots v_{i_n}=w_{i_1}\cdots w_{i_n}.$$
Construct the following qfst with input alphabet
$\{1,\ldots,k\}$: it has states $q_0,q_v,q_w$, and $q_{\rm rej}$.
The initial transformation produces a superposition of
$q_v,q_w,q_{\rm rej}$, each with amplitude $1/\sqrt{3}$. The unitaries
$U_i$ are all identity, but the output function is defined as
$f_i(q_x)=x_i$, for $x\in\{v,w\}$. The endmarker maps $q_v,q_w$ to accepting
states. It is clear that $i_1,\ldots,i_n$ is a solution iff
$(i_1\ldots i_n,v_{i_1}\cdots v_{i_n})$ is in the relation computed
with probability $2/3$ (the automaton is easily modified so that
it rejects when the input was the empty word, in this way we
force $n>0$).
\par
By replacing the unitaries by stochastic matrices (with entries
the squared moduli of the corresponding amplitudes) the same applies
to pfst.
\par
Since it is well known that PCP is undecidable, it follows that
there can be no decision procedure for emptiness of the relation
computed by the constructed pfst, or qfst, respectively.
\qed
\begin{rem}
  Undecidable questions for quantum finite automata were noted first
  for ``$1\frac{1}{2}$--way'' automata, i.e.
  ones which move only to the right on their input, but may also
  keep their position on the tape. In~\cite{amano:iwama} it is
  shown that the equivalence problem for these is undecidable.
  The same was proved for 1--way--2--tape quantum finite automata
  in~\cite{b:f:g}.
\end{rem}
\begin{conj}
  \label{conj:decidable}
  The emptiness problem for probabilistic and quantum fst computing a relation
  with probability $0.99$ is decidable.
  \par
  The emptiness problem for probabilistic and quantum fst computing a relation
  with a single--letter input alphabet, with probability $1/2+\varepsilon$ is decidable.
\end{conj}
To prove this, we would like to apply a packing argument in the space of
all total states, equipped with the above metric. However, this fails
because of the infinite volume of this space (for finite automata it
is finite, see~\cite{rabin} and~\cite{kondacs:watrous}). In any case,
a proof must involve the size of the gap between the upper and the lower
probability point, as the above theorem shows that it cannot possibly
work with gap $1/3$.
\par
Still, we can prove:
%
\begin{thm}
  If the relation ${\cal R}$ is computed by a pfst or a qfst with
  an isolated cutpoint, then
  ${\rm Range}({\cal R})=\{y:\exists x\ (x,y)\in{\cal R}\}$
  is a recursive set (so, for each specific output, it is decidable
  if it is ever produced above the threshold probability).
\end{thm}
{\it Proof}:
  Let the cutpoint be $\alpha$, with isolation radius $\delta$, and
  let $y=y_1\ldots y_n\in\Sigma_2^*$.\par
  Define
  $Y=\{y_1\ldots y_i:0\leq i\leq n\}$, the set of prefixes of $y$.
  Consider the \emph{output--$y$--truncated total state}, which is an
  element
  \begin{equation*}\begin{split}
    (\ket{\widetilde{\psi}},\widetilde{P}_{\rm ACC},\widetilde{p}_{\rm rej})
           &\in       \ell^2(Q\times Y)\oplus \ell^1(Y)\oplus \ell^1(\{{\rm REJ}\})      \\
           &\subseteq \ell^2(Q\times\Sigma_2^*)\oplus \ell^1(\Sigma_2^*)\oplus \ell^1(\{{\rm REJ}\}).
  \end{split}\end{equation*}
  It is obtained from $(\ket{\psi},P_{\rm ACC},p_{\rm rej})$
  -- with $\ket{\psi}=\sum_{q,w} \alpha_{qw}\ket{q}\otimes\ket{w}$ --
  by defining
  \begin{align*}
    \ket{\widetilde{\psi}}  &=\sum_{q\in Q,w\in Y} \alpha_{qw}\ket{q}\otimes\ket{w}, \\
    \widetilde{P}_{\rm ACC} &=P_{\rm ACC}|_{Y},                                      \\
    \widetilde{p}_{\rm rej} &=p_{\rm rej}+\sum_{q\in Q,w\not\in Y} |\alpha_{qw}|^2
                                         +\sum_{w\not\in Y} P_{\rm ACC}(w).
  \end{align*}
  Let us denote this transformation by $J$.
  Now observe that in the total state evolution of the qfst probability
  once put outside $Y$ never returns, and likewise, amplitude once
  put outside $Q\times Y$ never returns (compare proof of
  theorem~\ref{thm:qfa-c-qfst}). Formally, this is reflected in the
  relation
  $$J T_{ab}(\ket{\widetilde{\psi}},\widetilde{P}_{\rm ACC},\widetilde{p}_{\rm rej})
   =J T_b J T_a(\ket{\widetilde{\psi}},\widetilde{P}_{\rm ACC},\widetilde{p}_{\rm rej}).$$
  Hence, if we want to know if $T(y|x)\geq\alpha+\delta$ for some $x$,
  we may concentrate on the space of output--$y$--truncated total states,
  which is finite dimensional, and its transformation functions
  $\widetilde{T}_a=JT_a$.
  \par
  It is easily seen that there is a constant $\gamma$ such that
  for all truncated total states $s,t$ and all $w\in\Sigma_1^*$
  $$\|\widetilde{T}_w s-\widetilde{T}_w t\|\leq \gamma\|s-t\|.$$
  Hence, for $x,x',w\in\Sigma_1^*$, if
  $$\|\widetilde{T}_{\ddag x}(\ket{q_0}\otimes\ket{\epsilon},{\bf 0},0)
   -\widetilde{T}_{\ddag x'}(\ket{q_0}\otimes\ket{\epsilon},{\bf 0},0)\|<\delta/\gamma,$$
  then
  $$\|\widetilde{T}_{\ddag xw\$}(\ket{q_0}\otimes\ket{\epsilon},{\bf 0},0)
   -\widetilde{T}_{\ddag x'w\$}(\ket{q_0}\otimes\ket{\epsilon},{\bf 0},0)\|<\delta.$$
  Because of the cutpoint isolation we find that either
  both or none of $(x,y)$, $(x',y)$ is in ${\cal R}$.
  Now, because of compactness of the set of truncated total states
  reachable from the starting state, it follows that there is a constant
  $c>1$ such that for all $x\in\Sigma_1^*$ of length $|x|\geq c$
  one can write $x=v x_0 w$, with $0<|x_0|<c$, such that
  $$\|\widetilde{T}_{\ddag v x_0}(\ket{q_0}\otimes\ket{\epsilon},{\bf 0},0)
   -\widetilde{T}_{\ddag v}(\ket{q_0}\otimes\ket{\epsilon},{\bf 0},0)\|<\delta/\gamma.$$
  Hence 
  $$\|\widetilde{T}_{\ddag x\$}(\ket{q_0}\otimes\ket{\epsilon},{\bf 0},0)
   -\widetilde{T}_{\ddag vw\$}(\ket{q_0}\otimes\ket{\epsilon},{\bf 0},0)\|<\delta,$$
  and thus, if $x$ had produced $y$ with probability at least $\alpha+\delta$,
  so had the shorter string $vw$. This means that we only have to consider
  input strings of length up to $c$ to decide whether $y\in{\rm Range}({\cal R})$.
  \par
  Obviously, this reasoning applies to pfst, too.
\qed\par

\section{Deterministic vs. Probabilistic Transducers}
\label{sec:det:prob}
Unlike the situation for finite automata, pfst are strictly more powerful than
their deterministic counterparts:
\begin{thm}
  \label{thm:mmm}
  For arbitrary $\varepsilon>0$ the relation
  $${\cal R}_1=\{(0^m1^m,2^m):m\geq 0\}$$
  can be computed by a pfst with
  probability $1-\varepsilon$.
  It cannot be computed by a dfst.
\end{thm}
{\it Proof}:
  The idea is essentially from~\cite{frei:1}: for a natural number $k$
  choose initially an alternative $j\in\{0,\ldots,k-1\}$, uniformly.
  Then do the following: repeatedly read $k$ $0$'s, and output
  $j$ $2$'s, until the $1$'s start (remember the remainder modulo $k$), then
  repeatedly read $k$ $1$'s, and output $k-j$ $2$'s. Compare the remainder modulo
  $k$ with what you remembered: if the two are equal, output this number of
  $2$'s and accept, otherwise reject.
  \par
  It is immediate that on input $0^m1^m$ this machine outputs $2^m$ with
  certainty. However, on input $0^m1^{m'}$ each $2^n$ receives
  probability at most $1/k$.
  \par
  That this cannot be done deterministically is straightforward: assume that
  a dfst has produced $f(m)$ $2$'s after having read $m$ $0$'s. Because of
  finiteness there are $k,l$ such that after reading $k$ $1$'s (while $n_0$
  $2$'s were output) the internal state is the same as after reading $l$
  further $1$'s (while $n$ $2$'s are output). So, the output for input
  $0^m1^{k+rl}$ is $2^{f(m)+n_0+rn}$, and these pairs are either all
  accepted or all rejected. Hence they are all rejected, contradicting
  acceptance for $m=k+rl$.
\qed\par
By observing that the random choice at the beginning
can be mimicked quantumly, and that all intermediate computations 
are in fact reversible, we immediately get
\begin{thm}
  \label{thm:mmm:q}
  For arbitrary $\varepsilon>0$ the relation
  ${\cal R}_1$ can be computed by a qfst with
  probability $1-\varepsilon$. \qed
\end{thm}
Note that this puts qfst in contrast to quantum finite automata:
in~\cite{a:f} it was shown that if a language is recognized with probability
strictly exceeding $7/9$ then it is possible to accept it with
probability $1$, i.e. reversibly deterministically.
\begin{thm}
  The relation
  $${\cal R}_2=\{(w2w,w):w\in\{0,1\}^*\}$$
  can be computed by a pfst and by a qfst with probability $2/3$.
\end{thm}
{\it Proof}:
  We do this only for qfst (the pfst is obtained by replacing the unitaries
  involved by the stochastic matrices obtained by computing the squared
  moduli of the entries): let the input be
  $x2y$ (other forms are rejected).
  With amplitude $1/\sqrt{3}$ each go to one of three `subprograms':
  either copy $x$ to the output, or $y$ (and accept),
  or reject without output. This works by the same reasoning as the proof
  of theorem~\ref{thm:empty1}
\qed\par

\section{{...} vs. Quantum Transducers}
\label{sec:qfst}
After seeing a few examples one might wonder if everything that
can be done by a qfst can be done by a pfst. That this is not so
is shown as follows:
\begin{thm}
  \label{thm:mnk:q}
  The relation
  $${\cal R}_3=\{(0^m1^n2^k,3^m): n\neq k \wedge (m=k \vee m=n)\}$$
  can be computed by a qfst with probability $4/7-\varepsilon$, for
  arbitrary $\varepsilon>0$.
\end{thm}
\begin{thm}
  \label{thm:mnk}
  The relation ${\cal R}_3$ cannot be computed
  by a pfst with probability bounded away from $1/2$.
  In fact, not even with an isolated cutpoint.
\end{thm}
\par\noindent
{\it Proof (of theorem~\ref{thm:mnk:q})}:
  For a natural number $l$ construct the following transducer: from
  $q_0$ go to one of the states $q_1$, $q_{j,b}$ ($j\in\{0,\ldots,l-1\}$, $b\in\{1,2\}$),
  with amplitude $\sqrt{3/7}$ for $q_1$ and with amplitude
  $\sqrt{2/(7l)}$ each, for the others. Then proceed as follows (we assume the
  form of the input to be $0^m1^n2^k$, others are rejected):
  for $q_1$ output one $3$ for each $0$, and finally accept.
  For $q_{j,b}$ repeatedly
  read $l$ $0$'s and output $j$ $3$'s (remember the remainder $m\mod l$).
  Then repeatedly read $l$ $b$'s and output $l-j$ $3$'s (output nothing
  on the $(3-b)$'s). Compare the remainder with the one remembered, and reject
  if they are unequal, otherwise output this number of $3$'s.
  Reading $\$ $ perform the following unitary on the subspace
  spanned by the $q_{j,b}$ and duplicte states $q_{j',b}$:
  \begin{equation*}
    (j\leftrightarrow j')\otimes\frac{1}{\sqrt{2}}\left(\begin{array}{rr}
                                                          1 &  1 \\
                                                          1 & -1
                                                        \end{array}\right).
  \end{equation*}
  Accepting are all $q_{j',2}$, rejecting are all $q_{j',1}$.
  \par
  Now assume that the input does not occur as the left member in the relation:
  this means either $m\neq k$ and $m\neq n$, or $m=n=k$. In the first case
  all the outputs in each of the $b$--branches of the program
  are of different length, so get amplitude $\sqrt{2/(7l)}$. The final
  step combines at most two of them, so any output is accepted with
  probability at most $4/(7l)$. The second case is more interesting:
  in all branches the amplitude is concentrated on the output $3^m$.
  The rotation $V_\$ $ however is made such that the amplitude on
  $q_{j',2}$ cancels out, so we end up in a rejecting state $q_{j',1}$.
  In total, any output is accepted with probability at most $3/7+\varepsilon$.
  \par
  On the other hand, if the input occurs as the left member in the relation,
  exactly one of the two $b$--branches of the program concentrates all amplitude
  on output $3^m$, whereas the other spreads it to $l$ different lengths.
  This means that the output $3^m$ is accepted with probability
  at least $(l-1)\cdot 1/(7l)$, and others are accepted with probability
  at most $1/(7l)$ each.
  In total, the output $3^m$ is accepted with probability at least
  $4/7-\varepsilon$, all others are accepted with probability at most
  $3/7+\varepsilon$.
\qed\par\medskip
\noindent
{\it Proof (of theorem~\ref{thm:mnk})}:
  By contradiction.
  Suppose ${\cal R}_3$ is computed by a pfst $T$ with isolated cutpoint $\alpha$.
  The following construction computes it with probability bounded away from
  $1/2$: assuming $\alpha\leq 1/2$ (the other case is similar),
  let $p=\frac{1/2-\alpha}{1-\alpha}$. Run one of the following subprograms
  probabilistically: with probability $p$ output one $3$ for each $0$, and ignore
  the other symbols (we may assume that the input has the form $0^m1^n2^k$),
  with probability $1-p$ run $T$ on the input. It is easily seen that this
  new pfst computes the same relation with probability bounded away from $1/2$.
  \par
  Hence, we may assume that $T$ computes ${\cal R}$ with probability
  $\varphi>1/2$, from this we shall derive a contradiction.
  The state set $Q$ together with any of the stochastic matrices
  $V_0,V_1,V_2$ is a Markov chain. We shall use the classification of
  states for finite Markov chains (see~\cite{kemeny:snell}): for $V_i$
  $Q$ is partitioned into the set $R_i$ of \emph{transient} states
  (i.e. the probability to find the process in $R_i$
  tends to $0$) and a number of sets $S_{ij}$ of \emph{ergodic} states
  (i.e. once in $S_{ij}$ the process does not leave this set, and all
  states inside can be reached from each other, though maybe only by
  a number of steps). Each $S_{ij}$ is divided further into its
  \emph{cyclic} classes $C_{ij\nu}$ ($\nu\in\Z_{d_{ij}}$), $V_i$
  mapping $C_{ij\nu}$ into $C_{ij\nu+1}$. By considering
  sufficiently high powers $V_i^d$ (e.g. product of all the periods
  $d_{ij}$) as transition matrices, all these cyclic sets become ergodic,
  in fact, $V_i^d$ restricted to each is \emph{regular}.
  \par
  Using only these powers amounts to concentrating on input
  of the form ${\bf 0}^m{\bf 1}^n{\bf 2}^k$, with ${\bf i}=i^d$,
  which we will do from now on. Relabelling, the ergodic sets of
  $V_{\bf i}=V_i^d$ will be denoted $S_{ij}$. Each has its
  unique equilibrium distribution, to which every initial one
  converges: denote it by $\pi_{ij}$. Furthermore, there are
  limit probabilities $a(j_0)$ to find the process $V_{\bf 0}$
  in $S_{0j_0}$ after long time, starting from $q_0$.
  Likewise, there are limit probabilities $b(j_1|j_0)$ to
  find the process $V_{\bf 1}$ in $S_{1j_1}$ after long time,
  starting from $\pi_{0j_0}$, and similarly $c(j_2|j_1)$.
  So, by the law of large numbers, for large enough $m,n,k$
  the probability that $V_{\bf 0}$ has passed into $S_{0j_0}$
  after $\sqrt{m}$ steps, after which $V_{\bf 1}$ has passed into $S_{1j_1}$
  after $\sqrt{n}$ steps, after which $V_{\bf 2}$ has passed into $S_{2j_2}$
  after $\sqrt{k}$ steps, is arbitrarily close to
  $P(j_0,j_1,j_2)=a(j_0)b(j_1|j_0)c(j_2|j_1)$. (Note that these probabilities
  sum to one).
  \par
  As a consequence of the ergodic theorem (or law of large numbers),
  see~\cite{kemeny:snell},~ch.~4.2, in each of these events
  $J=(j_0,j_1,j_2)$ the probable number of $3$'s written after
  the final $\$ $, is linear in $m,n,k$:
  $$T(3^{[(1-\delta)\lambda_J(m,n,k),(1+\delta)\lambda_J(m,n,k)]}
                                |{\bf 0}^m{\bf 1}^n{\bf 2}^k,J)\rightarrow 1,$$
  as $m,n,k\rightarrow\infty$, with
  $$\lambda_J(m,n,k)=\alpha_J m+\beta_J n+\gamma_J k,$$
  and non--negative constants $\alpha_J,\beta_J,\gamma_J$.
  \par
  Since we require that for $k\neq m$
  $$T(3^{dm}|{\bf 0}^m{\bf 1}^m{\bf 2}^k)\geq\varphi,$$
  it is necessary that for a set ${\cal A}$ of events
  $J=(j_0,j_1,j_2)$
  $$\alpha_J+\beta_J=d,\ \gamma_J=0,\text{ with }P({\cal A})\geq \varphi.$$
  In fact, as for $J\not\in{\cal A}$
  $$T(3^{dm}|{\bf 0}^m{\bf 1}^m{\bf 2}^k,J)\rightarrow 0$$
  for certain sequences $m,k\rightarrow\infty$, we even have
  $$\sum_{J\in{\cal A}} P(J)T(3^{dm}|{\bf 0}^m{\bf 1}^m{\bf 2}^k,J)\geq\varphi-o(1).$$
  \par
  For $J\in{\cal A}$ it is obvious that the transducer
  outputs \emph{no more} $3$'s, once
  in $S_{2j_2}$. But this implies that for $m,k$ large enough,
  $T(3^{dm}|{\bf 0}^m{\bf 1}^m{\bf 2}^k,J)$ is arbitrarily close
  to $T(3^{dm}|{\bf 0}^m{\bf 1}^m{\bf 2}^m,J)$, hence
  $$T(3^{dm}|{\bf 0}^m{\bf 1}^m{\bf 2}^m)\geq \varphi-o(1),$$
  which implies that
  $$T(3^{dm}|{\bf 0}^m{\bf 1}^m{\bf 2}^m)\geq \varphi,$$
  contradicting $(0^{dm}1^{dm}2^{dm},3^{dm})\not\in{\cal R}_3$.
\qed\par
In general however, computing with isolated cutpoint is strictly weaker than
with probability bounded away from $1/2$ (observe that for finite automata,
probabilistic and quantum, recognizability
with an isolated cutpoint is equivalent to recognizability with probability
bounded away from $1/2$, see theorem~\ref{thm:qfa-c-qfst}):
\begin{thm}
  \label{thm:cutpoint}
  The relation
  $${\cal R}_4=\{(0^m1^na,4^l):(a=2 \rightarrow l=m) \wedge (a=3 \rightarrow l=n)\}$$
  can be computed by a pfst and by a qfst with an isolated cutpoint
  (in fact, one arbitrarily close to $1/2$),
  but not with a probability bounded away from $1/2$.
\end{thm}
\noindent
{\it Proof}:
  First the construction (again, only for qfst):
  initially branch into two possibilities
  $c_0,c_1$, each with amplitude $1/\sqrt{2}$. Assume that the
  input is of the correct form (otherwise reject), and in state
  $c_i$ output one $4$ for each $i$, ignoring the $(1-i)$'s.
  Then, if $a=2+i$, accept, if $a=3-i$, reject.
  It is easily seen that $4^l$ is accepted with probability
  $1/2$ if $(0^m1^na,4^l)\in{\cal R}_4$, and with probability
  $0$ otherwise.
  \par
  That this cannot be done with probability above $1/2$ is clear
  intuitively: the machine has to produce some output (because of
  memory limitations), but whether to output $4^m$ or $4^n$
  it cannot decide until seeing the last symbol.
  Formally, assume that $|m-n|>4t$, with
  $t=\max_{a,q} |f_a(q)|$. If
  $$T_{\ddag 0^m1^n2 \$}((q_0,\epsilon),{\bf 0},0)[4^m]=
                                  T(4^m|0^m1^n2)\geq 1/2+\delta,$$
  necessarily
  $$T_{\ddag 0^m1^n}((q_0,\epsilon),{\bf 0},0)[4^m]+
    T_{\ddag 0^m1^n}((q_0,\epsilon),{\bf 0},0)[Q_{\rm non}\times 4^{[m-2t,m+2t]}]
                                                \geq 1/2+\delta.$$
  But this implies
  $$T_{\ddag 0^m1^n}((q_0,\epsilon),{\bf 0},0)[4^n]+
    T_{\ddag 0^m1^n}((q_0,\epsilon),{\bf 0},0)[Q_{\rm non}\times 4^{[n-2t,n+2t]}]
                                                \leq 1/2-\delta,$$
  hence
  $$T_{\ddag 0^m1^n3 \$}((q_0,\epsilon),{\bf 0},0)[4^n]=
                                  T(4^n|0^m1^n3)\leq 1/2-\delta,$$
  contradicting $(0^m1^n3,4^n)\in{\cal R}_4$.
\qed\par

To conclude from these examples, however, that quantum is even better
than probabilistic, would be premature:
\begin{thm}
  \label{thm:lastsymbol}
  The relation
  $${\cal R}_5=\{(wx,x):w\in\{0,1\}^*,x\in\{0,1\}\}$$
  cannot be computed by a qfst with an isolated cutpoint.
  (Obviously it is computed by a pfst with probability $1$,
  i.e. a dfst).
\end{thm}
{\it Proof}:
  The construction of a dfst computing the relation is straightforward.
  To show that no qfst doing this job exists, we recall from~\cite{kondacs:watrous}
  that $\{0,1\}^*0$ is not recognized by a $1$--way quantum finite automaton
  with probability bounded away from $1/2$, and use theorem~\ref{thm:qfa-c-qfst}
  for this language.
\qed

\section{Conclusion}
\label{sec:concl}
We introduced quantum finite state transducers, and showed some of their
unique properties: undecidability of the emptiness problem, as opposed to
deterministic finite state transducers and finite automata, and incomparability
of their power to that of probabilistic and deterministic finite state transducers.
As open questions we would like to point out primarily our
conjecture~\ref{conj:decidable}. Another interesting question is whether a relation
computed by a qfst with probability \emph{sufficiently close to $1$}
can be computed by a pfst.
This would be the closest possible analog to the ``$7/9$--theorem'' from~\cite{a:f}.

\section*{Acknowledgements}
Research of RF supported by contract IST--1999--11234 (QAIP) from the European Commission,
and grant no. 96.0282 from the Latvian Council of Science. This work was carried out during
RF's stay at Bielefeld University in October 2000.
At this time AW was at Fakult\"at f\"ur Mathematik, Universit\"at Bielefeld,
Germany, and was supported by SFB 343 ``Diskrete Strukturen in der Mathematik''
of the Deutsche Forschungsgemeinschaft.


\end{document}